\documentclass[reprint,superscriptaddress,amsmath,amssymb,aps,prc,twocolumn,floatfix,nofootinbib]{revtex4-2}

\usepackage{graphicx}
\usepackage{hyperref}
\usepackage{xcolor}
\usepackage[english]{babel}
\usepackage{bm}
\usepackage[T1]{fontenc}
\usepackage{multirow}
\usepackage{booktabs}
\usepackage[utf8]{inputenc}

\usepackage[output-decimal-marker={.},exponent-product=\cdot]{siunitx}

\begin{document}

\title{Probing the quantum phase transition around ${N\approx60}$ via mass measurements of technetium isotopes}

\author{J. Ruotsalainen}
\thanks{These authors contributed equally}
\email{jouni.k.a.ruotsalainen@jyu.fi}
\affiliation{University of Jyväskylä, Department of Physics, Accelerator Laboratory, P.O. Box 35(YFL) FI-40014 University of Jyväskylä, Finland}
\author{A.~Jaries}
\thanks{These authors contributed equally}
\email{arthur.a.jaries@jyu.fi}
\affiliation{University of Jyväskylä, Department of Physics, Accelerator Laboratory, P.O. Box 35(YFL) FI-40014 University of Jyväskylä, Finland}
\author{M.~Stryjczyk}
\thanks{These authors contributed equally}
\email{stryjczyk@ill.fr}
\affiliation{University of Jyväskylä, Department of Physics, Accelerator Laboratory, P.O. Box 35(YFL) FI-40014 University of Jyväskylä, Finland}
\affiliation{Institut Laue-Langevin, 71 Avenue des Martyrs, F-38042 Grenoble, France}
\author{A.~Kankainen}
\email{anu.kankainen@jyu.fi}
\affiliation{University of Jyväskylä, Department of Physics, Accelerator Laboratory, P.O. Box 35(YFL) FI-40014 University of Jyväskylä, Finland}
\author{B. Andel}
\affiliation{Department of Nuclear Physics and Biophysics, Comenius University in Bratislava, 84248 Bratislava, Slovakia}
\author{M.~Araszkiewicz}
\affiliation{Faculty of Physics, University of Warsaw, PL 02-093 Warsaw, Poland}
\affiliation{National Centre for Nuclear Research (NCBJ), Andrzeja So\l{}tana 7, Otwock-\'Swierk, 05-400, Poland}
\author{O. Beliuskina}
\affiliation{University of Jyväskylä, Department of Physics, Accelerator Laboratory, P.O. Box 35(YFL) FI-40014 University of Jyväskylä, Finland}
\author{A.M.~Bruce}
\affiliation{School of Architecture, Technology and Engineering, University of Brighton, Brighton BN2 4GJ, United Kingdom}
\author{S. Cannarozzo}
\affiliation{Department of Physics and Astronomy, Uppsala University, Uppsala 75120, Sweden}
\author{S. Chinthakayala}
\affiliation{University of Jyväskylä, Department of Physics, Accelerator Laboratory, P.O. Box 35(YFL) FI-40014 University of Jyväskylä, Finland}
\affiliation{GANIL, CEA/DSM-CNRS/IN2P3, Boulevard Henri Becquerel, 14000 Caen, France}
\author{S. Doshi}
\affiliation{School of Architecture, Technology and Engineering, University of Brighton, Brighton BN2 4GJ, United Kingdom}
\author{T.~Eronen}
\affiliation{University of Jyväskylä, Department of Physics, Accelerator Laboratory, P.O. Box 35(YFL) FI-40014 University of Jyväskylä, Finland}
\author{A. Fija\l{}kowska}
\affiliation{Faculty of Physics, University of Warsaw, PL 02-093 Warsaw, Poland}
\author{L.M. Fraile}
\affiliation{Grupo de Física Nuclear \& IPARCOS, Universidad Complutense de Madrid, CEI Moncloa, 28040 Madrid, Spain}
\author{P. Garczy\'{n}ski}
\affiliation{Faculty of Physics, University of Warsaw, PL 02-093 Warsaw, Poland}
\author{Z.~Ge}
\affiliation{University of Jyväskylä, Department of Physics, Accelerator Laboratory, P.O. Box 35(YFL) FI-40014 University of Jyväskylä, Finland}
\author{D.~Grigorova}
\affiliation{Faculty of Physics, Sofia University "St. Kliment Ohridski", 1164 Sofia, Bulgaria}
\author{G. Jaworski}
\affiliation{Heavy Ion Laboratory, University of Warsaw, PL-02-093 Warsaw, Poland}
\author{A. Korgul}
\affiliation{Faculty of Physics, University of Warsaw, PL 02-093 Warsaw, Poland}
\author{T.~Krakowski}
\affiliation{National Centre for Nuclear Research (NCBJ), Andrzeja So\l{}tana 7, Otwock-\'Swierk, 05-400, Poland}
\author{J.~Kurpeta}
\affiliation{Faculty of Physics, University of Warsaw, PL 02-093 Warsaw, Poland}
\author{S. Lalkovski}
\affiliation{Faculty of Physics, Sofia University "St. Kliment Ohridski", 1164 Sofia, Bulgaria}
\author{M. Llanos-Exp\'{o}sito}
\affiliation{Grupo de Física Nuclear \& IPARCOS, Universidad Complutense de Madrid, CEI Moncloa, 28040 Madrid, Spain}
\author{I.D.~Moore}
\affiliation{University of Jyväskylä, Department of Physics, Accelerator Laboratory, P.O. Box 35(YFL) FI-40014 University of Jyväskylä, Finland}
\author{L.M. Motilla}
\affiliation{University of Jyväskylä, Department of Physics, Accelerator Laboratory, P.O. Box 35(YFL) FI-40014 University of Jyväskylä, Finland}
\affiliation{GANIL, CEA/DSM-CNRS/IN2P3, Boulevard Henri Becquerel, 14000 Caen, France}
\author{M.~Mougeot}
\affiliation{University of Jyväskylä, Department of Physics, Accelerator Laboratory, P.O. Box 35(YFL) FI-40014 University of Jyväskylä, Finland}
\author{H.~Penttil\"a}
\affiliation{University of Jyväskylä, Department of Physics, Accelerator Laboratory, P.O. Box 35(YFL) FI-40014 University of Jyväskylä, Finland}
\author{A.~Raggio}
\affiliation{University of Jyväskylä, Department of Physics, Accelerator Laboratory, P.O. Box 35(YFL) FI-40014 University of Jyväskylä, Finland}
\author{W. Rattanasakuldilok}
\affiliation{University of Jyväskylä, Department of Physics, Accelerator Laboratory, P.O. Box 35(YFL) FI-40014 University of Jyväskylä, Finland}
\author{J. Saren}
\affiliation{University of Jyväskylä, Department of Physics, Accelerator Laboratory, P.O. Box 35(YFL) FI-40014 University of Jyväskylä, Finland}
\author{K. Solak}
\affiliation{Faculty of Physics, University of Warsaw, PL 02-093 Warsaw, Poland}

\begin{abstract}
The masses of neutron-rich $^{104-106}$Tc isotopes were measured using the JYFLTRAP double Penning trap and found to deviate from the Atomic Mass Evaluation 2020 by $-79(25)$, $40(12)$ and $94(41)$~keV, respectively. In the case of $^{105,106}$Tc, the updated $Q_\beta$ values are in agreement with a previous JYFLTRAP measurement, disagreeing with the values from the mass evaluation. The new mass values result in a more linear trend in two-neutron separation energies indicating that technetium ($Z=43$) isotopes around $N \approx 60$ are not a part of the island of shape coexistence around $^{100}$Zr$_{60}$.
\end{abstract}

\maketitle

\section{Introduction}

Technetium (${Z=43}$) is the lightest element which has no stable isotopes. Due to its refractory properties, the technetium beams are not easily available at isotope separator on-line (ISOL) facilities. Consequently, only a limited number of studies on technetium isotopes have been conducted so far. The magnetic moments are known for $^{93-99}$Tc \cite{Walchli1952,Hinfurtner1995,Raeder2017,Stone2019,Kron2020} while quadrupole moments and isotope shifts, which are directly proportional to differences in mean-square charge radii, are measured only for the three longest-lived $^{97-99}$Tc isotopes \cite{Buttgenbach1982,Raeder2017,Kron2020}. The masses of the very neutron-rich $^{106-114}$Tc and very neutron-deficient $^{87-93,95}$Tc isotopes have been measured with Penning traps \cite{Hager2007,Weber2008,Hakala2011,Fallis2011,Haettner2011,Kankainen2012,Vilen2019,Ge2024}. However, for the less exotic $^{94,96-105}$Tc species, the masses are extracted from the transfer-reaction and decay studies \cite{Huang2021} which are prone to inaccuracies \cite{Goncharov2011,Nesterenko2019,Ge2021,Gamage2022,Hukkanen2023,Ge2023,Jaries2023}. Additionally, in the case of $^{106}$Tc the Penning-trap result was rejected from the Atomic Mass Evaluation 2020 (AME20) due to the presence of unidentified contaminants \cite{Huang2021}. 

The neutron-rich isotopes of technetium around ${N\approx60}$ are interesting from the nuclear structure point of view. In the region around $^{100}$Zr$_{60}$ sudden changes of multiple observables, such as two-neutron separation energies $S_{2n}$, differences in mean-square charge radii, or the $2^+_1$ excitation energies, have been observed, see Refs.~\cite{Heyde2011,Garrett2022} and references therein. These changes, called the quantum phase transition \cite{Togashi2016}, seem to reach up to the molybdenum isotopic chain at ${Z=42}$. However, since the masses of technetium isotopes in this region are known with a relatively low precision, more precise measurements are needed to accurately pinpoint the $S_{2n}$ trend. 

In addition to the fundamental understanding of nuclear structure, the $^{104-106}$Tc isotopes have been indicated to be some of the most impactful isotopes on the emission of the decay heat from fission of $^{239}$Pu \cite{Algora2024,Algora2010}. Decays of these species have been shown to have an effect on the so-called reactor anomaly, which is a discrepancy between the nuclear reactor antineutrino spectra observed experimentally and simulated based on the known decay data \cite{Fallot2012,Estienne2014}. Currently, to better understand the decay heat and the reactor anomaly, the main focus is on Total Absorption Spectroscopy (TAS) studies which allow for extraction of mean $\beta$ and $\gamma$ energies released in one decay event. At the same time, the decay energies ($Q_\beta$), which are deemed to be ‘‘key nuclear parameters in quantifying irradiated fuel inventories to determine total decay heat’’, are not expected to change \cite{Nichols2023}. However, since the $Q_\beta$ values are directly proportional to atomic masses, any changes in the latter will result in changes in the former, influencing our understanding of the energy release in time.

In this work we report on the precise mass measurements of $^{104-106}$Tc isotopes measured with the JYFLTRAP double Penning trap \cite{Eronen2012}. The implications of the results are discussed in the context of nuclear structure as well as data consistency.

\section{Experimental method}

Two experiments were performed at the Ion Guide Isotope Separator On-Line facility (IGISOL) \cite{Moore2013} using the JYFLTRAP double Penning trap mass spectrometer \cite{Kolhinen2004,Eronen2012} at the University of Jyväskylä, Finland. The masses of the $^{104,106}$Tc isotopes were measured during the first experimental run while $^{105}$Tc in the second one. In both campaigns, the technetium ions were produced in proton-induced fission of $^\text{nat}$U by irradiating a 15 mg/cm$^2$ target with 25-MeV protons from the K-130 cyclotron. The fission products were stopped and thermalized in a helium-filled gas cell, from which they were extracted by the helium flow and the electric field created by a radiofrequency sextupole ion guide \cite{Karvonen2008}. Then, the singly-charged Tc$^+$ ions were accelerated to 30 keV and transported to the IGISOL dipole magnet with a mass resolving power sufficient for further transferring isobarically pure beams. The mass-separated continuous ion beam was sent to the radiofrequency cooler-buncher \cite{Nieminen2001,Virtanen2025}, where the ions were thermalized and delivered as ion bunches to the JYFLTRAP Penning trap for further purification and mass measurements.

The JYFLTRAP consists of two cylindrical Penning traps inside a 7-T superconducting solenoid. The ion bunch is first injected into the helium-filled purification trap, where the buffer-gas cooling technique \cite{Savard1991} was used to select the desired isotope with a mass resolving power of $\approx 10^5$. The cooled and centered ions of interest were finally sent to the measurement trap for the mass measurements.

With a Penning trap, determining the mass of an ion is based on measuring its free space cyclotron frequency
\begin{equation}
    \nu_c=\frac{1}{2\pi}\frac{q}{m}B,
\end{equation}
where $q$ and $m$ are respectively the charge and mass of the ion while $B$ is the magnetic field strength in the trap. In this work, the cyclotron frequencies of the $^{104,106}$Tc were determined using the Phase-Imaging Ion Cyclotron Resonance (PI-ICR) method \cite{eliseev2013phase,Eliseev2014,Nesterenko2018,Nesterenko2021}. For $^{105}$Tc, the Time-of-Flight Ion Cyclotron Resonance (ToF-ICR) \cite{Konig1995} technique was used.

In both methods, by measuring the ratio of the cyclotron frequencies ${r=\nu_{c,\text{ref}}/\nu_{c}}$ of the ion-of-interest ($\nu_c$) and a reference  with a well-known mass ($\nu_{c,\text{ref}}$), both singly charged, the atomic mass of the nuclide of interest $m$ was obtained as
\begin{equation}
    m=r(m_{ref}-m_e)+m_e,
\end{equation}
where $m_{ref}$ is the atomic mass of the reference and $m_{e}$ the mass of the electron. A surface ion source \cite{Vilen2020} was utilized to produce the reference ions, $^{133}$Cs$^+$ in the case of $^{104,106}$Tc and $^{85}$Rb$^+$ for $^{105}$Tc. 

\begin{figure}[h!t!b]
\includegraphics[width = 1.1\columnwidth]{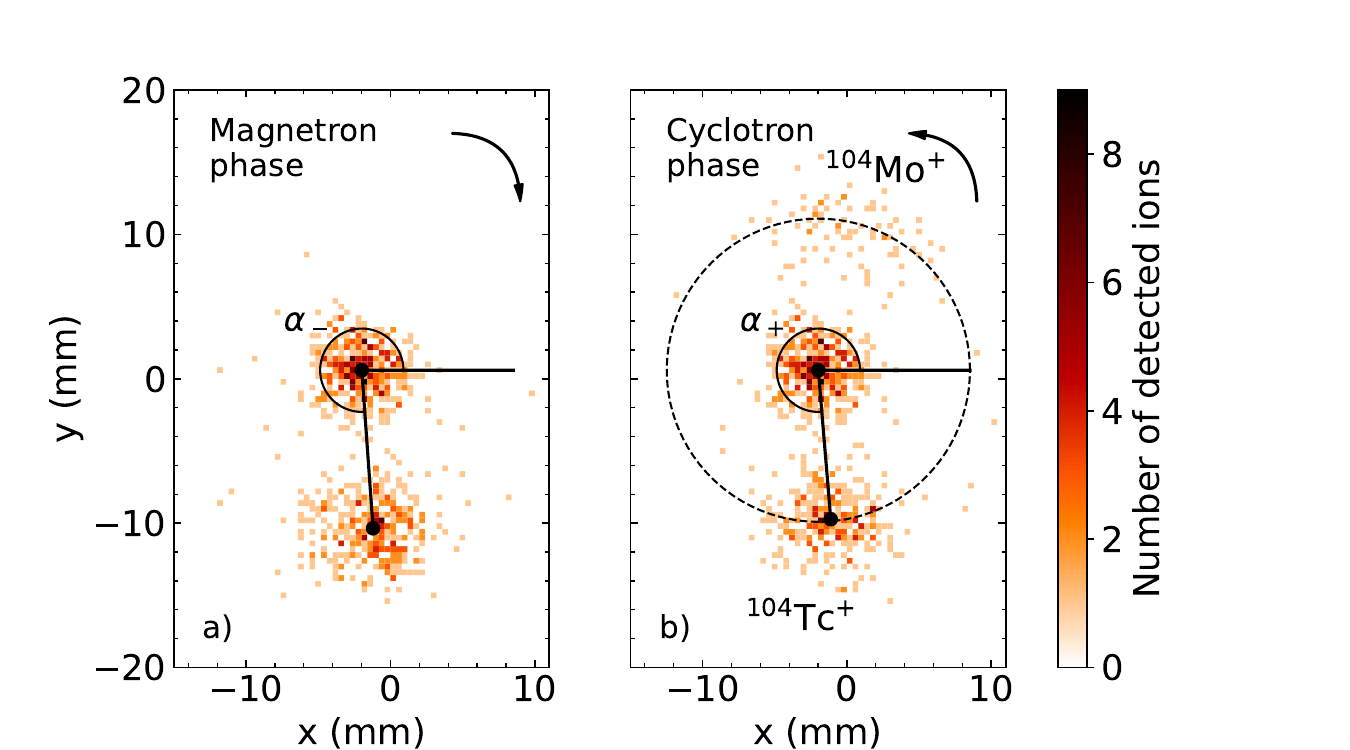}
\caption{\label{fig:piicr_tc104}Magnetron \textbf{a)} and cyclotron \textbf{b)} motion phase projections of $^{104}$Tc$^+$ ions on a 2D-MCP using the PI-ICR technique with $t_\text{acc} = 482$~ms, together with trap center spot. The isobaric contaminant $^{104}$Mo$^+$ is also identified. A dashed circle on \textbf{b)} indicates the average cyclotron excitation radius. The polar angles between the reference phase, the center of the trap, and the magnetron ($\alpha_-$ on \textbf{a)}) and cyclotron ($\alpha_+$ on \textbf{b)}) phase spots are marked with an arc.}
\end{figure}

\begin{figure}[h!t!b]
\includegraphics[width = 1.1\columnwidth]{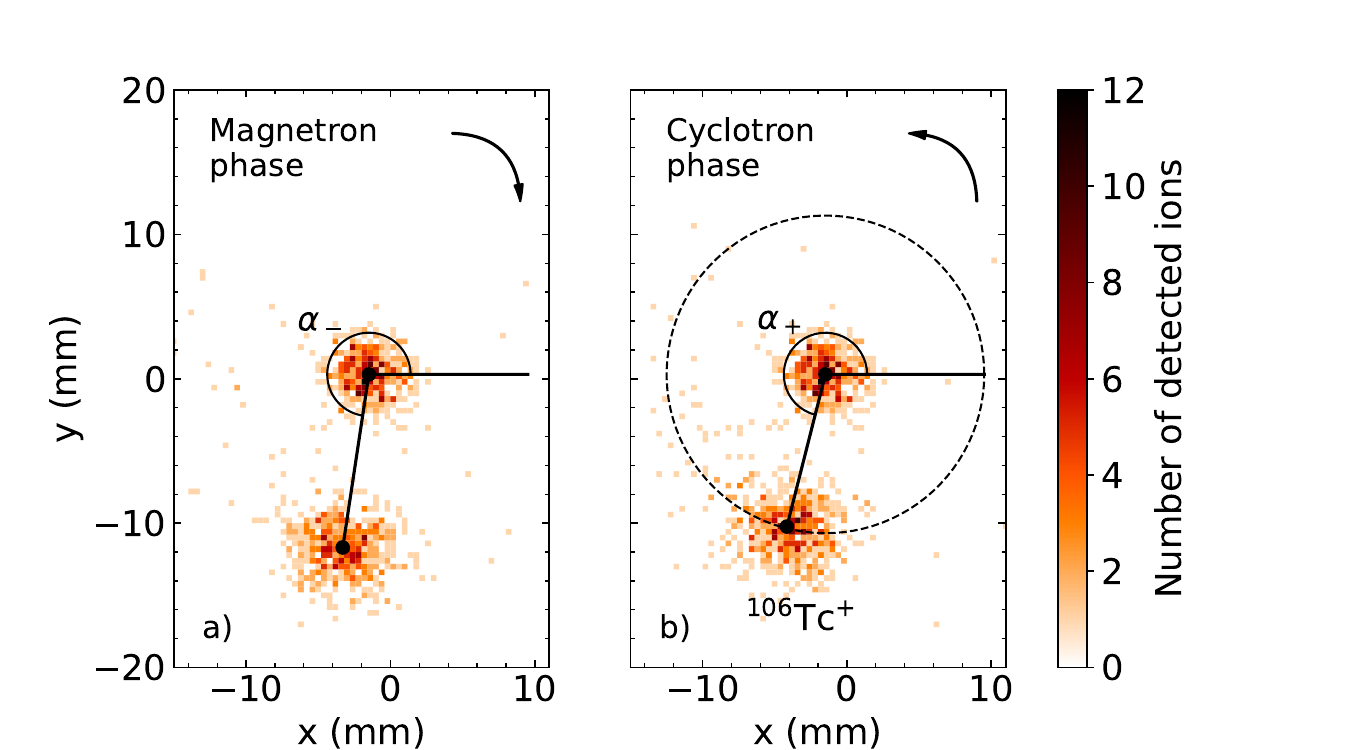}
\caption{\label{fig:piicr_tc106}Magnetron \textbf{a)} and cyclotron \textbf{b)} motion phase projections of $^{106}$Tc$^+$ ions on a 2D-MCP using the PI-ICR technique with $t_\text{acc} = 715$~ms, together with trap center spot. A dashed circle on \textbf{b)} indicates the average cyclotron excitation radius. The polar angles between the reference phase, the center of the trap, and the magnetron ($\alpha_-$ on \textbf{a)}) and cyclotron ($\alpha_+$ on \textbf{b)}) phase spots are marked with an arc.}
\end{figure}

In the first campaign, the PI-ICR technique was used to determine the $^{104,106}$Tc masses. The images of the ion's radial eigenmotions are detected using a position-sensitive microchannel plate detector (2D MCP). A phase accumulation time of 482~ms was applied for $^{104}$Tc, sufficient to resolve it from the isobaric contaminant $^{104}$Mo (see Fig.~\ref{fig:piicr_tc104}). For $^{106}$Tc, a longer phase accumulation time of 715 ms was utilized  (see Fig.~\ref{fig:piicr_tc106}). The frequency of the $^{133}$Cs reference was measured for 5 minutes before and after a measurement of the ion-of-interest to linearly interpolate the magnetic field strength to the time of its measurement. $^{104}$Tc was measured for 14 minutes and $^{106}$Tc for 20 minutes before switching back to the reference. Two frequency ratios were measured for $104$Tc and for $^{106}$Tc, with around $\approx$100 $^{104}$Tc and $\approx$200 $^{106}$Tc ions collected per ratio. 

\begin{figure}[h!t!b]
\includegraphics[width = 1.0\columnwidth]{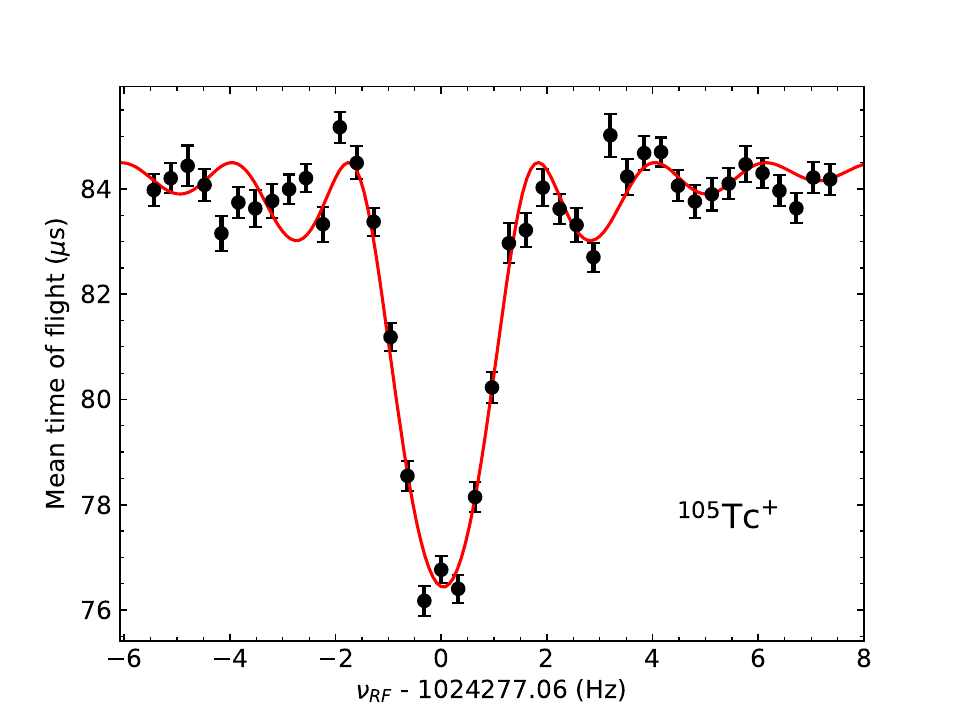}
\caption{\label{fig:toficr_105tc}Time-of-flight resonance for $^{105}$Tc$^+$ ions with a 500-ms excitation. The black data points indicate the mean time of flight of the ions for each scanned frequency $\nu_{RF}$. The solid red line is a fit of the theoretical lineshape \cite{Kretzschmar2007} to the data.}
\end{figure}

In the second campaign, the mass of $^{105}$Tc was determined using the ToF-ICR method to measure its cyclotron frequency $\nu_c$ (see Fig. \ref{fig:toficr_105tc}) with $^{85}$Rb used as a reference. Similarly as with the PI-ICR method, the cyclotron frequency measurements of the reference and ion-of-interest were interleaved to determine $B$, changing the species after 4 minutes of reference and 7 minutes of the ion of interest. Altogether three frequency ratios with $\approx$600 ions of $^{105}$Tc$^+$ each were obtained using a conversion time of 500 ms.

Ion bunches with up to three detected ions per bunch were taken into account in the analysis of $^{104,106}$Tc and up to five for the $^{105}$Tc measurement, as the performed count-rate class analysis \cite{Kellerbauer2003} did not show an effect on the measured frequency based on the number of ions. The uncertainty due to magnetic field fluctuation, measured to be $\delta B/(B\delta t) = 2.01\times10^{-12}$ /min \cite{Nesterenko2021}, was negligible compared to the statistical and other systematic uncertainties, as each measurement lasted for at most tens of minutes. Due to the mass difference between the ion of interest and the reference, a mass-dependent uncertainty of $\delta_m r/r = -2.35(81) \times 10^{-10} / \textnormal{u} \times (m_\text{ref} - m_\text{ion})$  \cite{Nesterenko2021} was quadratically added to the uncertainty of the ratios obtained for $^{104-106}$Tc and the ratios were also shifted by the same amount. After accounting for the mass-dependent shift, additional systematic effect has been seen to remain \cite{Nesterenko2021}. Thus a residual uncertainty of $\delta_\text{res}r/r=9\times 10^{-9}$ was quadratically added to the uncertainty of the ratios. For $^{104}$Tc, where isobaric contamination of $^{104}$Mo was identified, a correction for the magnetron phase advancement was applied \cite{Nesterenko2021}.  As several frequency ratios were obtained, the weighted average and the larger of the internal and external error \cite{Birge1932} were taken as the final ratio and its error. 

\section{Results and discussion}

\begin{table*}
\caption{\label{tab:results}The frequency ratios $r = \nu_{c,ref}/\nu_{c}$ and corresponding mass-excess values $\Delta$ determined in this work. Reference ions (Ref.), the excitation time in the ToF-ICR measurement of $^{105}$Tc and the phase accumulation time for the PI-ICR measurements of $^{104,106}$Tc are indicated with $t_\text{trap}$. The AME20 \cite{AME2020} values ($\Delta_\text{lit.}$) and differences between this work and the evaluation (${\mathrm{Diff.} = \Delta-\Delta_\text{lit.}}$) are also added. Half-lives ($T_{1/2}$) and spin-parity assignments ($J^{\pi}$) are taken from NUBASE20 \cite{NUBASE20}, with $\#$ indicating a non-experimental estimate.}
\begin{ruledtabular}
\begin{tabular}{lllllllll}
Nuclide     & $T_{1/2}$     & $J^{\pi}$ & Ref.  & $t_\text{trap}$ (ms)    & $r$   & $\Delta$ (keV)    & $\Delta_\text{lit.}$ (keV) & Diff. (keV)  \\\hline
$^{104}$Tc & 18.3(3) min    & ($3^-$)   & $^{133}$Cs &    482       &    \num{0.781843237(18)}   &     \num{-82578.3(22)}              & \num{-82499(25)}          & $-79(25)$\\
$^{105}$Tc & 7.64(6) min    & ($3/2^-$) & $^{85}$Rb &         500     &  \num{1.23553971(11)}     &  \num{-82196.1(86)}                 & \num{-82290(40)}          & $94(41)$\\
$^{106}$Tc & 35.6(6) s      & $(1,2)^{(+\#)}$ & $^{133}$Cs &  715      &   \num{0.796914549(15)}    & \num{-79736.1(18)}                  & \num{-79776(12)}          &  $40(12)$\\
\end{tabular}
\end{ruledtabular}
\end{table*}

The measured frequency ratios, the corresponding mass-excess values and the comparison to the Atomic Mass Evaluation 2020 (AME20) \cite{AME2020} are presented in Table~\ref{tab:results}. All the reported cases differ from the literature by up to 3.3 standard deviations. Since the literature (AME20) values are based on old beta-decay studies of $^{104-106}$Mo and $^{104-106}$Tc, our direct mass measurements indicate that dedicated ion trap experiments are required to achieve highly precise and accurate masses. We also note that the new mass excess of $^{106}$Tc, \num{-79736.1(18)}~keV, obtained with the PI-ICR technique, differs by $8.9(48)$~keV from the previous JYFLTRAP value, \num{-79745.0(44)}~keV\footnote{This is a reevaluated mass-excess value based on the frequency ratio reported in Ref. \cite{Hager2007} and the mass of the reference $^{102}$Ru from AME20 \cite{AME2020}.}, which was based on a ToF-ICR measurement reported in Ref. \cite{Hager2007}. The Penning-trap value from Ref.~\cite{Hager2007} was not fully utilized in the AME20 evaluation due to unidentified impurities in the trap. Here, with the PI-ICR technique, we are confident that there are no impurities in the measurement as shown in Fig.~\ref{fig:piicr_tc106}. 

\begin{table}
\caption{\label{tab:qbeta}$Q_\beta$ values of molybdenum and technetium isotopes re-evaluated using the $^{104-106}$Tc masses measured in this work compared with the AME20 values \cite{AME2020} and the JYFLTRAP results from Ref. \cite{Jordan2013}. In case of the $Q_\beta(^{105}\text{Mo})$ the mass of $^{105}$Mo from the recent JYFLTRAP measurement \cite{Jaries2024a} was taken.}
\begin{ruledtabular}
\begin{tabular}{llllll}
\multirow{2}{*}{$A$}     & \multicolumn{2}{c}{$Q_\beta$(Mo) (keV)}     & \multicolumn{3}{c}{$Q_\beta$(Tc) (keV)} \\\cmidrule(lr){2-3}\cmidrule(lr){4-6}
& This work & AME20 & This work & AME20 & Ref. \cite{Jordan2013} \\\hline
104 & 2234(9) & 2155(24) & 5517.5(33) & 5597(25) & 5516(6) \\
105 & 4863(9) & 4960(40) & 3738(9) & 3650(40) & 3746(6) \\
106 & 3608(9) & 3648(15) & 6578(5) & 6547(11) & - \\
\end{tabular}
\end{ruledtabular}
\end{table}

A comparison of the re-evaluated $Q_\beta$ values for $^{104-106}$Mo and $^{104-106}$Tc based on the technetium masses measured in this work and a previous JYFLTRAP measurement of $^{105}$Mo \cite{Jaries2024a}, previously measured $Q_\beta$ of $^{104,105}$Tc \cite{Jordan2013}, and the $Q_{\beta}$ values given in AME20 is presented in Table~\ref{tab:qbeta}. While the differences in masses between this work and AME20 result in discrepancies in the $Q_\beta$ values, the decay energies of $^{104,105}$Tc measured previously at JYFLTRAP, which are not included in the AME20 evaluation \cite{Huang2021}, agree perfectly with our current work. We note that the discrepancy between the $Q_\beta(^{105}\mathrm{Tc})$ value from AME2016 \cite{AME2016}, which is identical to AME20, and the previous measurement led the ENSDF evaluators to reject the two highest pseudolevels at 3660 and 3700~keV in $^{105}$Ru \cite{Lalkovski2019} although those were reported in Ref. \cite{Jordan2013}. Our new $Q_\beta(^{105}\mathrm{Tc})=3738(9)$~keV shows that beta decay can feed these pseudolevels and solves this issue.

\begin{figure}[h!t!b]
\centering
\includegraphics[width=\columnwidth]{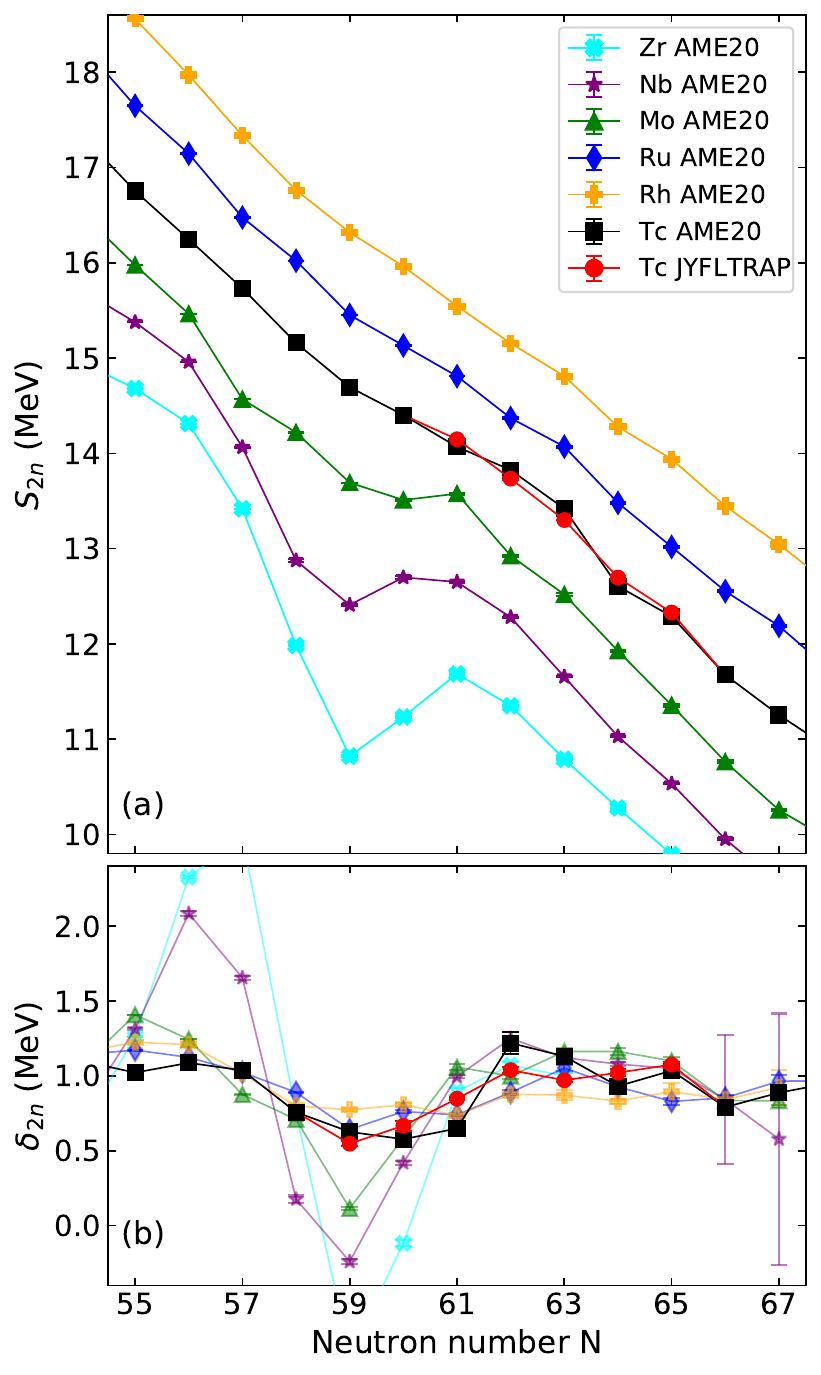}
\caption{\label{fig:masstrends}Mass trends, (a) two-neutron separation energy $S_{2n}$ and (b) two-neutron shell-gap energy $\delta_{2n}$, in the ${N\approx60}$ region plotted for isotopic chains from Zr ($Z=40$) to Rh ($Z=45$) based on the literature values from AME20 \cite{AME2020} and this work.}
\end{figure}

To check whether the sudden changes in the ground-state structure around ${N\approx60}$ are present in the technetium isotopic chain, the two-neutron separation energies $S_{2n}$:
\begin{equation}
S_{2n}(Z,N) = \Delta(Z,N-2) - \Delta(Z,N) +2\Delta_n \mathrm{,}
\end{equation}
were calculated. $\Delta(Z,N)$ corresponds to the mass excess of a nucleus with a given number of protons ($Z$) and neutrons ($N$) while $\Delta_n$ is the free neutron mass-excess value. Additionally, the two-neutron shell-gap energies $\delta_{2n}$ defined as:
\begin{equation}
\delta_{2n}(Z,N) = S_{2n}(Z,N) - S_{2n}(Z,N+2) \mathrm{,}
\end{equation}
were computed. The results from this work are compared with the values from AME20 for neighboring isotopic chains, from zirconium ($Z=40$) to rhodium ($Z=45$), and presented in Fig.~\ref{fig:masstrends}.

The updated $S_{2n}$ curve for the technetium isotope is flatter compared to the one calculated with the AME20 values. It is similar to the isotopic chains of ruthenium ($Z=44$) and rhodium ($Z=45$) rather than molybdenum ($Z=42$), niobium ($Z=41$) and zirconium ($Z=40$). This behavior is better visible when analyzing the $\delta_{2n}$ parameters. For the three latter elements a strong negative peak at $N=59$ is present, while for the two former and technetium the curves are rather smooth. Therefore, our measurements firmly confirm that the deviations of the $S_{2n}$ values from a linear trend observed around ${N\approx60}$ for the elements around zirconium are not present in the technetium isotopic chain.

\section{Summary and Outlook}

The high-precision mass measurements of the neutron-rich $^{104-106}$Tc isotopes were performed at the IGISOL facility using the JYFLTRAP double Penning trap. The new mass values were found to differ by up to 94 keV from the literature \cite{AME2020}. The two-neutron separation energies updated with the new values follow a rather linear trend, similar to the heavier isotopic chains of ruthenium and rhodium. This observation suggests that as $Z$ increases, the range of the island of shape coexistence around $^{100}$Zr extends to the molybdenum isotopes only. Our results also influenced the decay energies in the region and resolve a discrepancy between AME20 and a previous measurement \cite{Jordan2013} for the $Q_\beta$ values of $^{104,105}$Tc.

Our work clearly shows that masses of species which are close to the valley of stability should be reinvestigated with accurate experimental techniques, such as Penning traps. They have a significant influence on our understanding of the nuclear structure as well as for nuclear applications, such as the decay heat of nuclear reactors. Based on the TAS measurement priorities in Table 2 from Ref.~\cite{Nichols2023}, we propose that the isotopes $^{98,99}$Nb and $^{142}$La should be measured with the highest priority. $^{92-94}$Y, $^{101-103}$Tc, $^{129,130}$Sb, $^{132}$I, $^{139}$Cs, $^{141}$La and $^{145-147}$Pr are also of importance, as their masses have not been measured with a Penning trap. Our measurement of $^{106}$Tc, as well as the previous studies on $^{91}$Sr, $^{95}$Y and $^{138}$I \cite{Jaries2024,Beliuskina2024}, also indicate that the masses of reactor-relevant nuclei should be measured in multiple experiments to increase the trustworthiness of the evaluated results. 

\begin{acknowledgments}

This project has received funding from the European Union’s Horizon 2020 Research and Innovation Programme under Grant Agreements No. 771036 (ERC CoG MAIDEN) and No. 861198–LISA–H2020-MSCA-ITN-2019, from the European Union’s Horizon Europe Research and Innovation Programme under Grant Agreement No. 101057511 (EURO-LABS), the Next Generation EU, the National Recovery and Resilience Plan of the Republic of Bulgaria, project No. BG-RRP-2.004-0008-C01, from the Research Council of Finland projects No. 295207, No. 306980, No. 327629, No. 354589 and No. 354968, from the Polish National Science Center, Grant No. 2024/53/N/ST2/03168, No. 2020/39/B/ST2/02346, No. 2023/51/D/ST2/02816 and No. 2020/39/D/ST2/00466, from the Bulgarian National Science Fund contract number KP-06-N68/8, from the Slovak Research and Development Agency, Contract No. APVV-22-0282, the Slovak grant agency VEGA, Contract No. 1/0019/25 and from the UK Science and Technology Facilities Council, Grant No. ST/V001078. J.R. acknowledges financial support from the Vilho, Yrj\"o and Kalle V\"ais\"al\"a Foundation. J.K. acknowledges the support from the JYU Visiting Fellow Programme Grants 2022 and 2024.

\end{acknowledgments}

\bibliography{biblio}

\end{document}